# Defect engineering spin centers in interacting many-body Su-Schrieffer-Heeger chains

Lin Wang [*] and Thomas Luu [†]
*Institute for Advanced Simulation (IAS-4), Forschungszentrum Jülich, Germany*

Ulf-G. Meißner [‡]
*Helmholtz Institut für Strahlen- und Kernphysik and Bethe Center for Theoretical Physics; Universität Bonn, D-53115 Bonn, Germany; Institute for Advanced Simulation (IAS-4), Forschungszentrum Jülich, Germany; and Peng Huanwu Collaborative Center for Research and Education, International Institute for Interdisciplinary and Frontiers, Beihang University, Beijing 100191, China*

The ability to engineer topologically distinct materials opens the possibility of enabling novel phenomena in low-dimensional nanosystems, as well as manufacturing novel quantum devices. One of the simplest examples, the SSH model with both even and odd number of sites, demonstrates the connection between localized edge states and the topology of the system. We show that the SSH model hosts localized spin centers due to the interplay between the localized edge states and the on-site Hubbard interaction. We further show how one can engineer any number of localized spin centers within the chain by careful addition of defects. These spin centers are paired in spin-singlet or spin-triplet channels within each block separated by the defects, and together they construct an array of spin singlet and/or triplet qubits. The effect of on-site disorder on spin centers is also addressed. As this system is realizable experimentally, our findings describe a way for manipulating and engineering spin qubits and therefore provide a platform for performing many-body quantum simulations on spin excitations like magnons and triplons.

## I. INTRODUCTION

The Su-Schrieffer-Heeger (SSH) model [1] is the quintessential example demonstrating the connection between topology of a system and its localized edge states [2,3]. The model describes spinless electrons hopping via nearest neighbors on a one-dimensional chain. The hopping amplitudes take on two values, $t_1$ for "intracell" and $t_2$ for "intercell" hoppings as shown in Fig. 1(a). Depending on the ratio of these amplitudes, the system can reside in a nontrivial topological phase and thus support localized edge states.

In recent years, the SSH model has received more and more attention both theoretically and experimentally. Theoretically, the SSH model has been explored in the presence of on-site interactions [4], inter- and intracell spin-orbit couplings [5], superconducting pairing [6], chirality-preserving and chirality-breaking disorder [7–11], and nonlinearity induced by wavefunction amplitude dependent hoppings [12], for example. Progress has been equally impressive experimentally, where here the SSH model has been realized in various systems such as superconductors [13], LC waveguides [14], Rydberg atoms [15], micropillars [16], Hardcore bosons [17], nanoparticles [18], silicon quantum dots [19], artificial lattices made of Cs/InAs(111)A [20], trapped-ion chains controlled by local Floquet fields [21], and so on. Among these systems, the silicon quantum dots [19], artificial lattices made of Cs/InAs(111)A [20], and trapped-ion chains controlled by local Floquet fields [21] are of particular interest due to their ability to simulate the many-body SSH model by controlling the interaction strength as well as the intracell and intercell hoppings. A chain of quantum dots or artificial atoms is connected with tunable hopping parameters governed by the relative distances between intra- and intercells. In [19,20] it was demonstrated how such systems can be manipulated to exist in either the trivial or nontrivial phase even in the presence of strong interactions. This provides a highly controllable and useful platform for future simulations on topological properties and strong quantum correlations.

In addition to topological properties, the interactions may also influence other properties of the many-body SSH model especially in the nontrivial phase with localized edge states. The localization in the presence of strong interactions is expected to induce local magnetism. We show that both even and odd SSH chains with localized edge states can host localized spin centers at the ends supported by three different methods, i.e., exact diagonalization (ED) for small systems, full quantum Monto Carlo simulations (QMC), and mean-field (MF) theory calculations. Further, by careful inclusion of "defects" via modification of select hopping parameters

[*]Contact author: l.wang@fz-juelich.de
[†]Contact author: t.luu@fz-juelich.de
[‡]Contact author: meissner@hiskp.uni-bonn.de

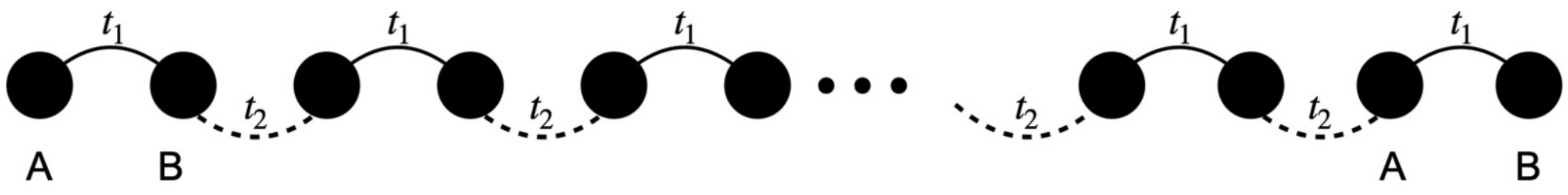

(a) Schematic picture of one-dimensional SSH model

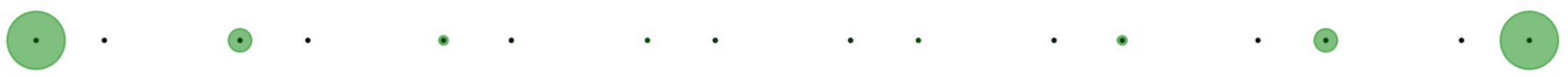

(b) Topological phase $t_1 < t_2$

FIG. 1. (a) A and B represent two sites in one unit cell. $t_1$ ($t_2$) denotes the intracell (intercell) hopping. (b) In the SSH model with an even number of sites in a topological phase $t_1 < t_2$, the probability density of the state with lowest positive energy is shown as green circles. Note that the SSH chain with an even number of sites ABAB $\cdots$ AB starts with A at the left side and ends with B at the right side.

within the chain, we can engineer more localized spin centers induced by these defects. These spin centers can be either in the spin-singlet or spin-triplet configuration within each block separated by the defects, and we find that the energy difference between such configurations is small compared to the single-particle gap. This is very different from previous work where defects were usually introduced in the noninteracting SSH model to investigate the robustness of topologically protected localized edge states against different types of disorders [7–11]. Therefore, our proposal provides a unique way to engineer spin centers in such chains by constructing an array of spin qubits consisting of spin singlets and/or spin triplets. These spin qubits can be used to simulate the dynamics of spin excitations like magnons and triplons [22].

Our paper is organized as follows. In the next section we review the topological properties of the noninteracting single-particle SSH model and describe how these properties translate to a finite system of even and odd sites. We then present our formalism to study the interacting many-body SSH model by including interactions with an onsite Hubbard term in Sec. III, via the mean-field theory calculations, full QMC simulations, as well as the ED method. Further, we introduce in Sec. IV the notion of a "defect" and demonstrate how such controllable defects can induce localized spin centers within the many-body SSH chain. In Sec. V we discuss the experimental feasibility of our work. Finally, we recapitulate and comment on potential applications of this work in Sec. VI.

## II. FORMALISM

As shown in Fig. 1(a), the SSH model is a one-dimensional lattice model with two sites A and B per unit cell. The tight-binding Hamiltonian with the nearest-neighbor hopping is given by

$$H_{\mathrm{SSH}} = \sum_i (t_1 c^\dagger_{i,\mathrm{A}} c_{i,\mathrm{B}} + t_2 c^\dagger_{i+1,\mathrm{A}} c_{i,\mathrm{B}} + \mathrm{H.c.}), \tag{1}$$

where $i$ represents the index of unit cell, $c^\dagger_{i,j}$ ($c_{i,j}$) are the creation (annihilation) operators of a particle on site $j = \mathrm{A/B}$ of the $i$th unit cell, and $t_1$ ($t_2$) denotes the intracell (intercell) hopping. The SSH model has sublattice symmetry or chiral symmetry with a unitary operator $C$ such that $\{H_{\mathrm{SSH}}, C\} = 0$. $C$ is a diagonal matrix with $+1(-1)$ for all A(B) sites. With chiral symmetry, the topological invariant is obtained by integrating the closed Berry connection over the 1D Brillouin zone, resulting in a $Z$ winding number [23,24]. As a result, the system can be divided into three distinct phases depending on the ratio between the hopping parameters: (i) $Z = 1$ topological (nontrivial) phase $t_1 < t_2$, (ii) $Z = 0$ trivial phase $t_1 > t_2$, and (iii) gapless phase $t_1 = t_2$.

### A. SSH chain with an even number of sites

To investigate the topological properties, i.e., bulk-edge correspondence of the SSH model, we calculate the eigenstates of the finite SSH chain with an even number of sites as shown in Fig. 1(b). In agreement with the nonzero topological invariant, two localized edge states with energies $E \sim 0$ are observed in the topological phase $t_1 < t_2$. The fact that the energies are not exactly zero arise from the unavoidable overlap between these two localized states. To reduce such overlap, the finite SSH chain has to be much longer than the decay length $\xi = \frac{1}{\ln(t_2/t_1)}$ of the localized edge states.[1]

### B. SSH chain with an odd number of sites

To overcome the disadvantages in the SSH chain with an even number of sites, we extend our investigation to the SSH chain with an odd number of sites. Compared with the even SSH model, the odd SSH model has one unpaired A or B site at the end. Here, we take the odd SSH model with one unpaired A, for example [see Fig. 2(a)]. For the odd SSH model, the existence of chiral symmetry guarantees at least one exactly zero-energy state, independent of the system size and also the choice of the hopping parameters [11]. To verify that these zero-energy states are localized, we show in Fig. 2(b) the probability density of the zero-energy state in the case of $t_1 < t_2$. We find that the zero-energy state is well localized at the left end. This can be understood as follows. The zero-energy states have well-defined chirality [25], either $+1$ or $-1$. For the odd SSH chain, only one of these two chiralities is allowed, i.e., $+1$ in our case. The chirality $+1$ requires the wavefunctions of all B sites to be exactly zero and the wavefuncitons of all adjacent A sites to be determined by the ratio of the hopping parameters. Specifically, when $t_1 < t_2$, zero-energy state has to be localized at the left side

[1]We provide a derivation of this decay length in Appendix A.

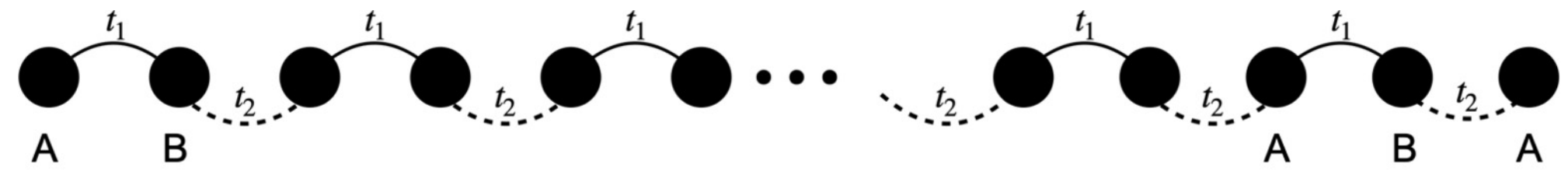

(a) Schematic picture of SSH model with odd number of sites

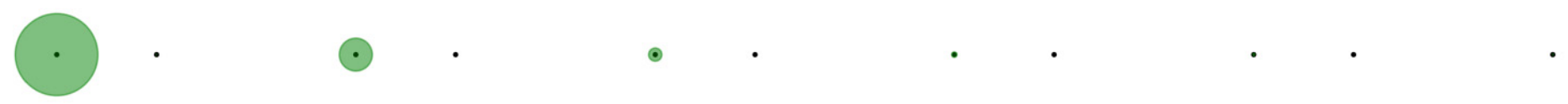

(b) $t_1 < t_2$

FIG. 2. (a) Odd SSH model with one unpaired A at the end. (b) In the SSH model with odd number of sites in the case of $t_1 < t_2$, the probability density of the state with exactly zero energy is plotted as green circles. Note that the SSH chain with an odd number of sites ABAB $\cdots$ ABA starts with A at the left side and ends with A at the right side.

with $\phi_{i,\text{A}} = (-t_1/t_2)^{i-1}\phi_{1,\text{A}}$, $\phi_{i,\text{A}}$ is the $i$th component of the A sites. More details are shown in Appendix A.

## III. MANY-BODY SSH CHAIN WITH HUBBARD INTERACTION

To introduce electron correlations we assign spin labels $\sigma = \uparrow, \downarrow$ to our electrons and include an onsite Hubbard interaction to (1),

$$H_{\text{SSH+U}} = \sum_{i,\sigma}(t_1 c^\dagger_{i,\sigma,\text{A}} c_{i,\sigma,\text{B}} + t_2 c^\dagger_{i+1,\sigma,\text{A}} c_{i,\sigma,\text{B}} + \text{H.c.}) - \frac{U}{2}\sum_x (n_{x,\uparrow} - n_{x,\downarrow})^2. \tag{2}$$

Here, $\sum_x$ represents the sum over all lattice sites, and $n_{x,\sigma} \equiv c^\dagger_{x,\sigma} c_{x,\sigma}$ is the electron number operator of spin $\sigma$ at site $x$. The form of this Hamiltonian ensures that the global ground state $|\Omega\rangle$ corresponds to the electrically neutral, "half-filled" state.

We are interested in calculating properties of the ground state $|\Omega\rangle$ of this system, such as the mean energy per site $\epsilon$,

$$\epsilon = \frac{1}{N_x}\langle\Omega|\hat{H}_{\text{SSH+U}}|\Omega\rangle, \tag{3}$$

with $N_x$ being the size of the system as well as the net spin $S_z$ and $S_z^2$ per site $x$,

$$\begin{aligned}\langle\Omega|\hat{S}_z(x)|\Omega\rangle &= \tfrac{1}{2}\langle\Omega|(n_{x,\uparrow} - n_{x,\downarrow})|\Omega\rangle \\ &= \tfrac{1}{2}(\langle n_{x,\uparrow}\rangle - \langle n_{x,\downarrow}\rangle) \equiv \frac{m_x}{2},\end{aligned} \tag{4}$$

$$\begin{aligned}\langle\Omega|\hat{S}_z^2(x)|\Omega\rangle &= \tfrac{1}{4}\langle\Omega|(n_{x,\uparrow} - n_{x,\downarrow})^2|\Omega\rangle \\ &= \tfrac{1}{4}(\langle n_{x,\uparrow}\rangle + \langle n_{x,\downarrow}\rangle - 2\langle n_{x,\uparrow} n_{x,\downarrow}\rangle).\end{aligned} \tag{5}$$

In this work we employ three different methods to calculate these quantities. First, if the system size is sufficiently small, we use ED to obtain these quantities exactly. Exact diagonalization also allows us access to the full many-body Hilbert space spectrum, providing us a means to calculate these quantities within a temperature bath,

$$\langle\hat{O}\rangle_\beta = \frac{1}{Z(\beta)}\sum_m \langle m|Oe^{-\beta E_m}|m\rangle; \quad Z(\beta) = \sum_m e^{-\beta E_m}. \tag{6}$$

Here $\beta$ is an inverse temperature, $Z$ is the partition function, and the sum is over all states $|m\rangle$ in the antisymmetric Fock space $F_-$ where

$$H_{\text{SSH+U}}|m\rangle = E_m|m\rangle \quad \forall\ |m\rangle \in F_-(\mathcal{H}),$$

and $\mathcal{H}$ represents the one-body Hilbert space. We apply this method to the systems with size of $N_x = 4$ and $N_x = 5$.

The second method we employ is MF theory. Here the mean-field approximation is applied to the number density operators,

$$n_{x,\sigma} n_{x,\sigma'} \approx \langle n_{x,\sigma}\rangle n_{x,\sigma'} + n_{x,\sigma}\langle n_{x,\sigma'}\rangle - \langle n_{x,\sigma}\rangle\langle n_{x,\sigma'}\rangle,$$

where $\langle n_{x,\sigma}\rangle = \sum_{E\leqslant E_f} |\psi_{E,\sigma}(x)|^2$, with $E_f$ the Fermi energy at half-filling. The wavefunctions $\psi_{E,\sigma}(x)$ themselves are obtained by diagonalizing the mean-field Hamiltonian,

$$H_{\text{mf}} = H_{\text{SSH}} - U\sum_x m_x(n_{x,\uparrow} - n_{x,\downarrow}) + \frac{U}{2}\sum_x m_x^2. \tag{7}$$

This formalism is solved self-consistently till a prescribed precision in $m_x$ is achieved.

Our final method involves calculating these quantities directly from QMC simulations. Here one performs stochastic estimates of Eq. (6) at a given inverse temperature $\beta$. This method has been documented in detail in, e.g., Refs. [26–28], and has been used to measure the nature of localized states in the presence of strong correlations [29,30]. We choose $\beta = 10$ with the number of timeslices $N_t = 80$ for our simulations. As our QMC simulations are fully nonperturbative they should agree (within statstics and discretization errors) with ED calculations if available. For large systems where ED results are not available, such as those we consider in Sec. IV, we consider the QMC results to be the standard. Note that the quantity $\langle\Omega|\hat{S}_z^2(x)|\Omega\rangle$ in all three methods is sufficient to show spin localizations. However, to probe specific spin configurations $\langle\Omega|\hat{S}_z(x)|\Omega\rangle$, only the MF method is possible by choosing proper initial conditions for $m_x$ that breaks the spin degeneracy of the ground state. Such information is not attainable from ED and QMC methods because the contribution to $\langle\Omega|\hat{S}_z(x)|\Omega\rangle$ from spin up and spin down exactly cancels each other at all positions. In principle, one could lift this spin degeneracy by including a small external staggered magnetic field, but unfortunately this is beyond the abilities of our QMC simulations at the moment.

We note that all numerical results quoted in this work are normalized by $\max(t_1, t_2)$ and thus dimensionless. For example, the mean energy, onsite coupling, and inverse temperature are "$\epsilon'' = \epsilon / \max(t_1, t_2)$, "$U'' = U/\max(t_1, t_2)$, and "$\beta'' = \beta \max(t_1, t_2)$, respectively. Finally, when plotting the locations of ions in our figures, we choose to represent their relative locations with simple integer indices. Their actual physical locations are system specific and depend on the type of experiment.

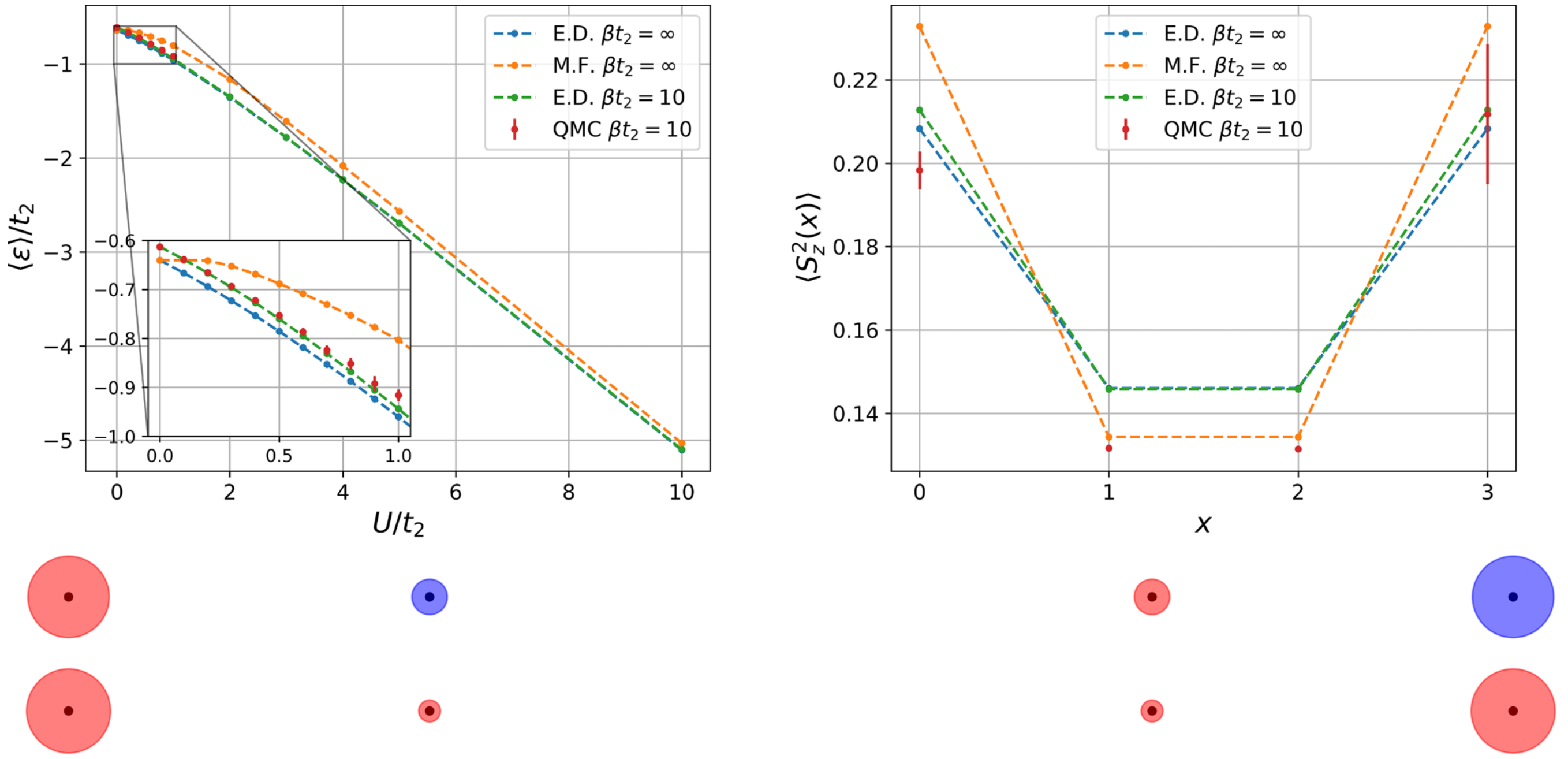

FIG. 3. Even SSH chain with four sites and $t_1/t_2 = 0.4$. The top plots show the comparison of select observables using exact diagonalization (ED) at zero temperature ($\beta = \infty$), ED with inverse temperature $\beta = 10$, mean field (MF) at zero temperature, and full quantum Monte Carlo (QMC) simulations with $\beta = 10$. (Top left) Average energy per site $\langle \epsilon \rangle$ over a range of Hubbard onsite coupling $U$; (top right) $\langle S_z^2(x) \rangle$ evaluated at $U = 0.8$. (Bottom) The spin density distribution of spin singlet pairing with $S_z = 0$ and spin triplet pairing with $S_z = 1$ obtained via MF at $U = 0.5$ is shown as circles where red and blue ones represent spin up and spin down, respectively. $S_z$ is the total spin.

In Fig. 3 we compare these methods for the $N_x = 4$ SSH model in the nontrivial phase with $t_1/t_2 = 0.4$. As expected there is excellent agreement between the finite-temperature ED results and the QMC results. When comparing the MF results at zero temperature with the corresponding ED results, both quantitative and qualitative agreements are shown at very large $U$. This is because the local fluctuations are extremely weak at very large $U$ even for small systems. When the interactions are very weak, a phase transition is predicted at critical Hubbard interaction $U_c^{mf} \sim 0.17$ using the MF method. However, this phase transition is not supported by the ED calculations since the MF method is not applicable in the presence of very strong local fluctuations. For the interactions in between within the range of $U > U_c^{mf}$, we find qualitative agreement in all three methods by considering both $\langle \epsilon \rangle$ and $\langle S_z^2(x)\rangle$. For the $N_x = 5$ SSH chain, we find quantitative agreement at very large $U$ and qualitative agreement at small $U$ on both the energy and $\langle S_z^2(x)\rangle$ in all three methods, as shown in Fig. 4.

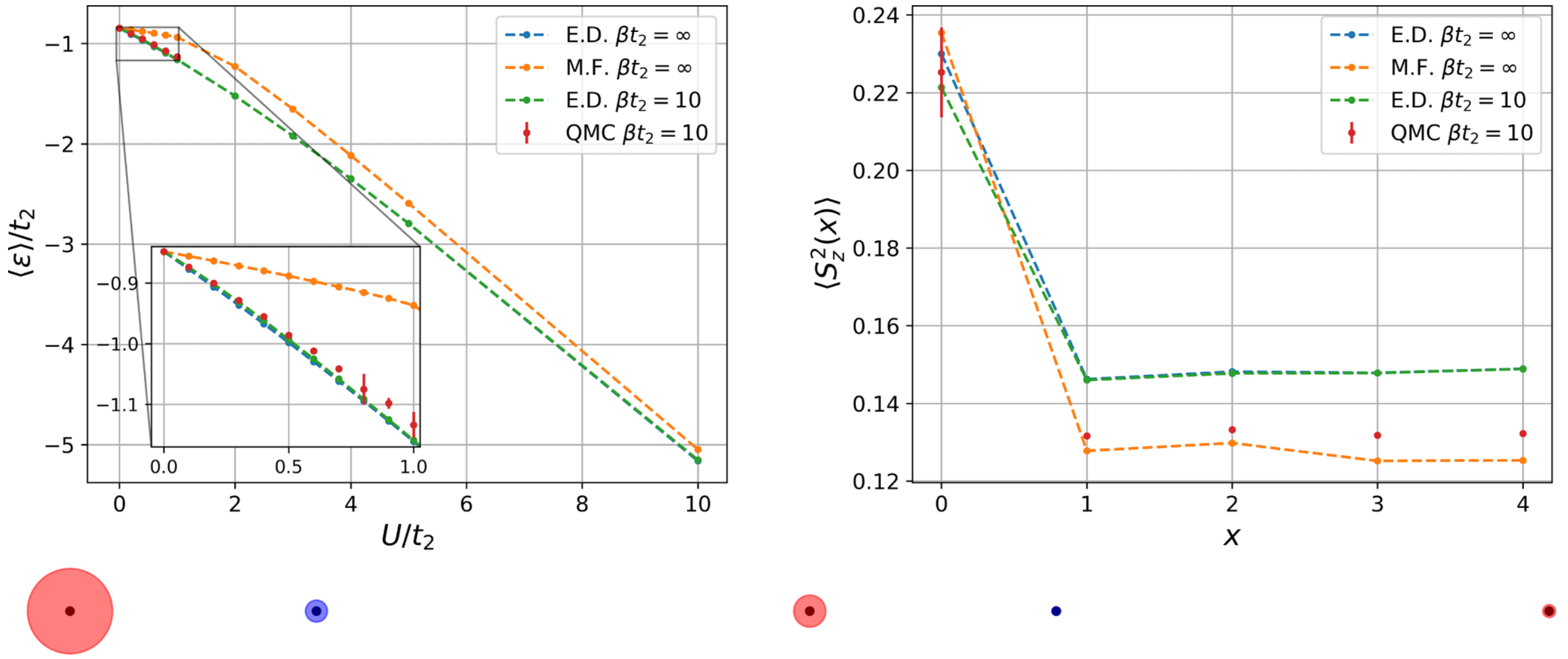

FIG. 4. Odd SSH chain with five sites and $t_1/t_2 = 0.4$. The top plots show the comparison of select observables using exact diagonalization (ED) at zero temperature ($\beta = \infty$), ED with inverse temperature $\beta = 10$, mean field (MF) at zero temperature, and full quantum Monte Carlo (QMC) simulations with $\beta = 10$. (Top left) average energy per site $\langle \epsilon \rangle$ over a range of Hubbard onsite coupling $U$; (top right) $\langle S_z^2(x) \rangle$ evaluated at $U = 0.8$. (Bottom) The spin density distribution of unpaired spin center at the left end with $S_z = 1/2$ obtained via MF at $U = 0.5$ is shown as circles where red and blue ones represent spin up and spin down, respectively. $S_z$ is the total spin.

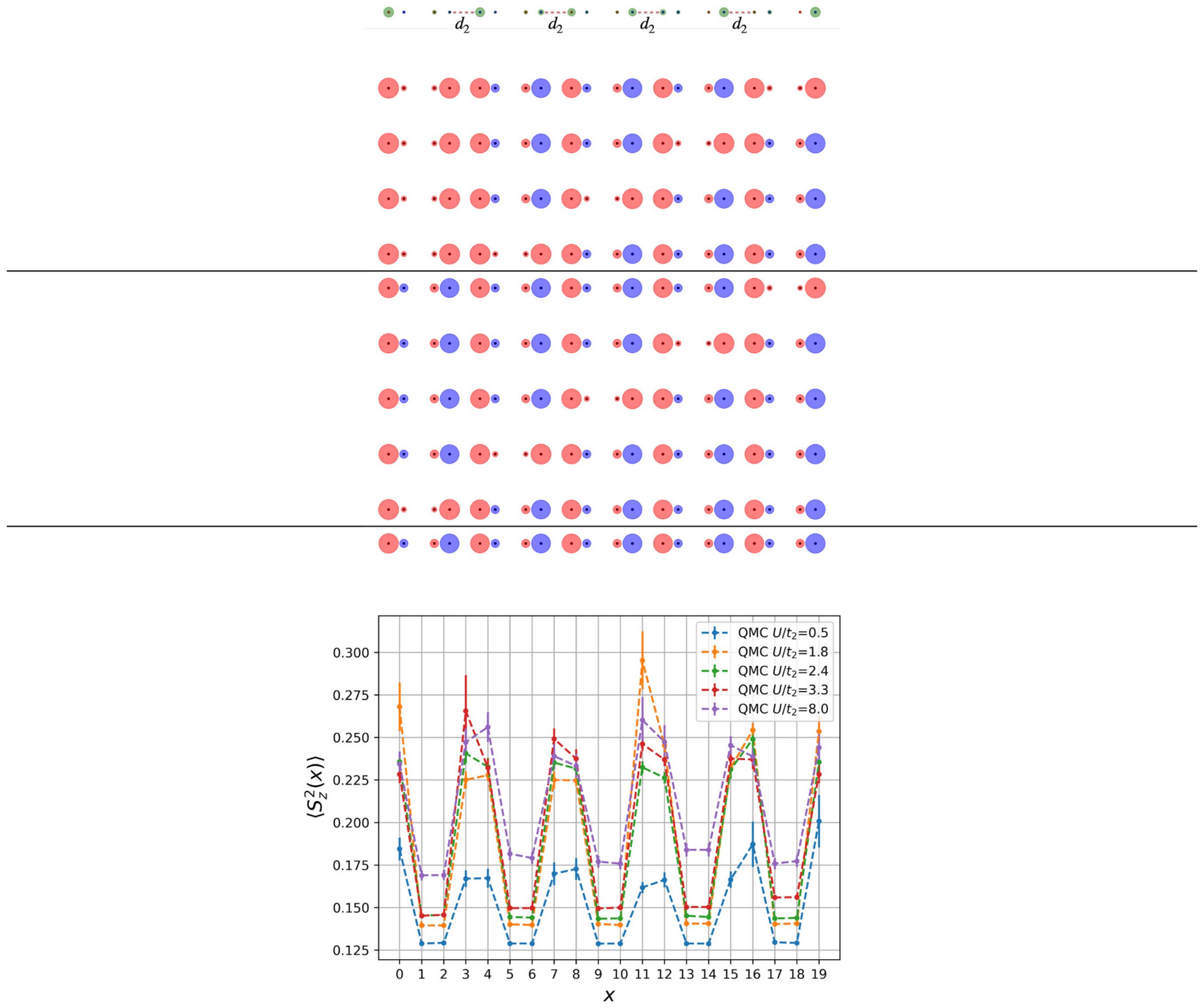

FIG. 5. Defect engineering the even SSH chain with $N_x = 20$, $t_1/t_2 = 0.4$, and Hubbard interaction $U$. The defects are introduced by modifying the hoppings $t_2 \to d_2$. (Top) The probability density of state with exactly zero energy at $U = 0$ is plotted as green circles. (Middle) Spin density distribution of the states with total spin $S_z = 0, 1, 2$ from bottom to top obtained via MF at $U = 0.5$ is shown as red (spin up) and blue (spin down) circles. (Bottom) $\langle S_z^2(x)\rangle$ of the ground state obtained from QMC results with select values of $U$ and $\beta = 10$. $S_z$ is the total spin. $d_2/t_2 = (t_1/t_2)^2$.

Despite the limitations of mean-field theory (e.g., it falsely predicts a phase transition at small $U$ in the even SSH chain), the MF method is still useful because it can show specific spin configurations when $U > U_c^{mf}$. For example, we show in the bottom of Figs. 3 and 4 specific spin density configurations (red corresponds to up, blue to down) obtained from the mean-field self-consistent solution. These localizations with large local spin densities at the end of the SSH chain are defined as spin centers. Spin centers are then paired on adjacent centers with singlet or triplet configurations at the two ends of the even SSH chain or unpaired at one end in the odd SSH chain. Such configurations depend on the initial conditions $m_x$ and provide us with a picture of the spin-energy landscape of states near the global ground state. As we show in the next section, this feature of MF will be helpful in understanding the role of spin localizations in the presence of defects.

These two examples corresponding to the even and odd SSH chains with paired and unpaired spin centers can be viewed as the basic building blocks from which to construct

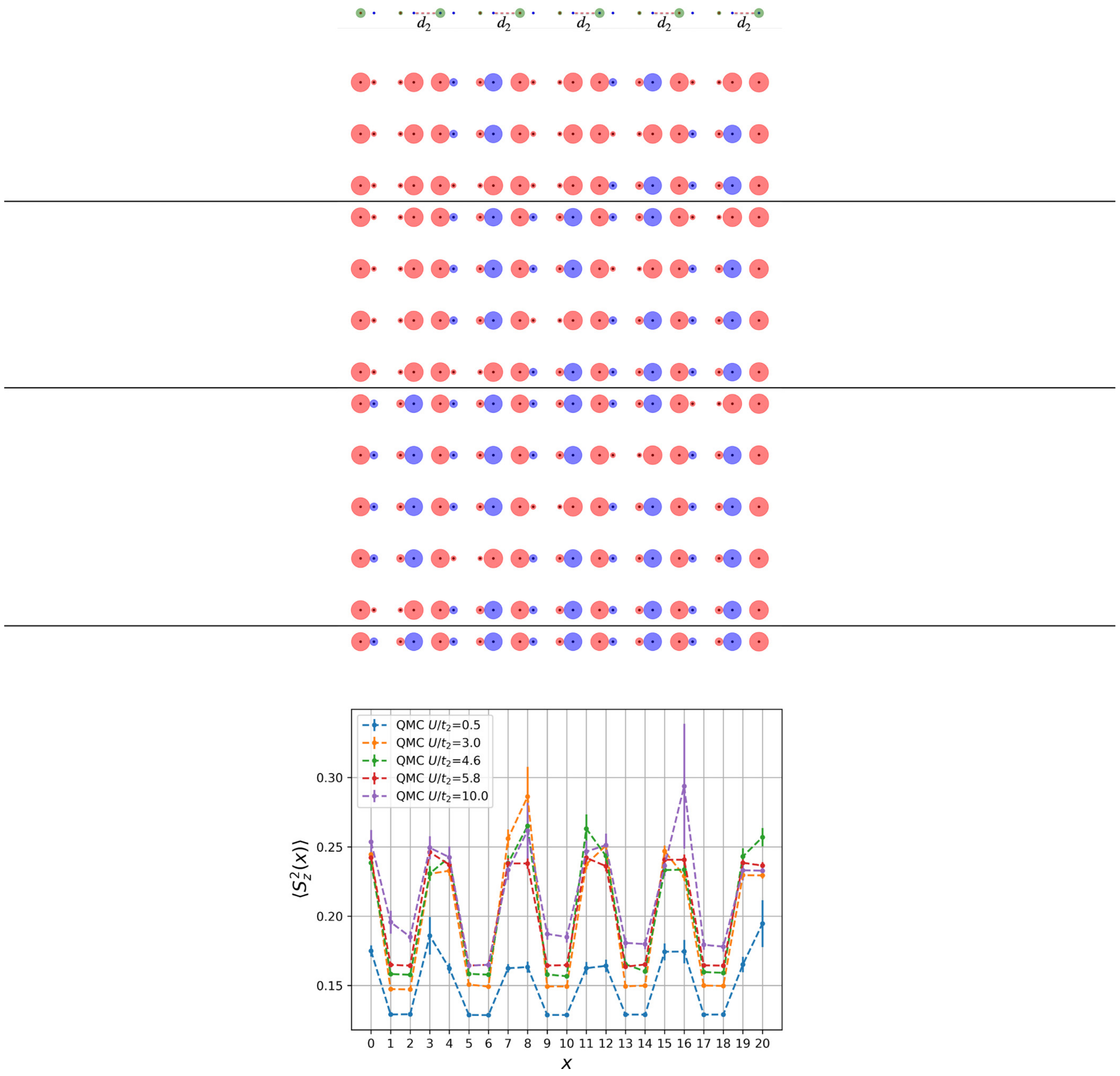

FIG. 6. Defect engineering the odd SSH chain with $N_x = 21$ and $t_1/t_2 = 0.4$ in the presence of Hubbard interaction $U$. The defects are introduced by modifying the hoppings $t_2 \to d_2$. (Top) The probability density of state with exactly zero energy at $U = 0$ is plotted as green circles. (Middle) Spin density distribution of the states with total spin $S_z = \frac{1}{2}, \frac{3}{2}, \frac{5}{2}, \frac{7}{2}$ from bottom to top obtained via MF at $U = 0.5$ is shown as red (spin up) and blue (spin down) circles. (Bottom) $\langle S_z^2(x)\rangle$ of the ground state obtained from QMC results with select values of $U$ and $\beta = 10$. $S_z$ is the total spin. $d_2/t_2 = (t_1/t_2)^2$.

longer SSH chains connected by defects, which we now consider in the following section.

## IV. USING DEFECTS TO ENGINEER SPIN CENTERS

In both even and odd SSH chains, paired or unpaired spin centers arise due to the joint effect of localized edge states and the on-site Hubbard interaction. Without interactions, on the other hand, there would be no spin centers. These spin centers can be used, for example, as spin qubits to form the building blocks of quantum simulation [31], quantum communication [32], and quantum sensing [33]. To design more spin centers, one simple way is to introduce controllable defects to engineer the magnetic properties of a finite SSH chain. Here, the defects are added by modifying the hoppings between two nearby sites. More details are given in Appendix B.

We start from the even SSH chain, e.g., with 20 sites the hopping ratio $t_1/t_2 = 0.4$, in the topological phase where spin singlet or triplet pairing at the two ends are expected. However, a single spin singlet or spin triplet is not enough for real applications. To generate more spin pairings with singlet and/or triplet configurations, we have to break the long SSH chain into several short blocks by defects. In the top of Fig. 5, the defects are introduced by changing the hoppings from $t_2$ to $d_2$ between the sites marked by the red dashed lines. These defects break the SSH chain into five blocks with a "4-4-4-4-4" configuration. Here, the spin centers are defined as the localizations with large local spin densities at the end of each block. With the Hubbard interaction $U = 0.5$ as shown in the middle of Fig. 5, the ground state with total spin zero consists of five paired spin singlets. All these spin centers are supported by the QMC calculations of $\langle S_z^2(x)\rangle$ shown in the bottom of Fig. 5. In addition, from bottom to top, we find that the excited states with larger total spins such as $S_z = 1$ and 2 host one, two, or more paired spin triplets to replace the spin singlet pairing in the ground state. Within each block with four sites, the pairing can be spin singlet or spin triplet as plotted in the bottom of Fig. 3. Therefore, the even SSH chain with defect engineering can be used to realize a system with total integer spin hosting a number of connecting spin singlet and/or triplet pairings. More configurations using defect engineering such as 18 sites with a "6-6-6" configuration and 20 sites with nonequal "6-4-6-4" configuration are shown in Figs. 9 and 10 in the Appendix. Note that the even SSH chain in the trivial phase is not a promising candidate for defect engineering due to the absence of localized edge states.

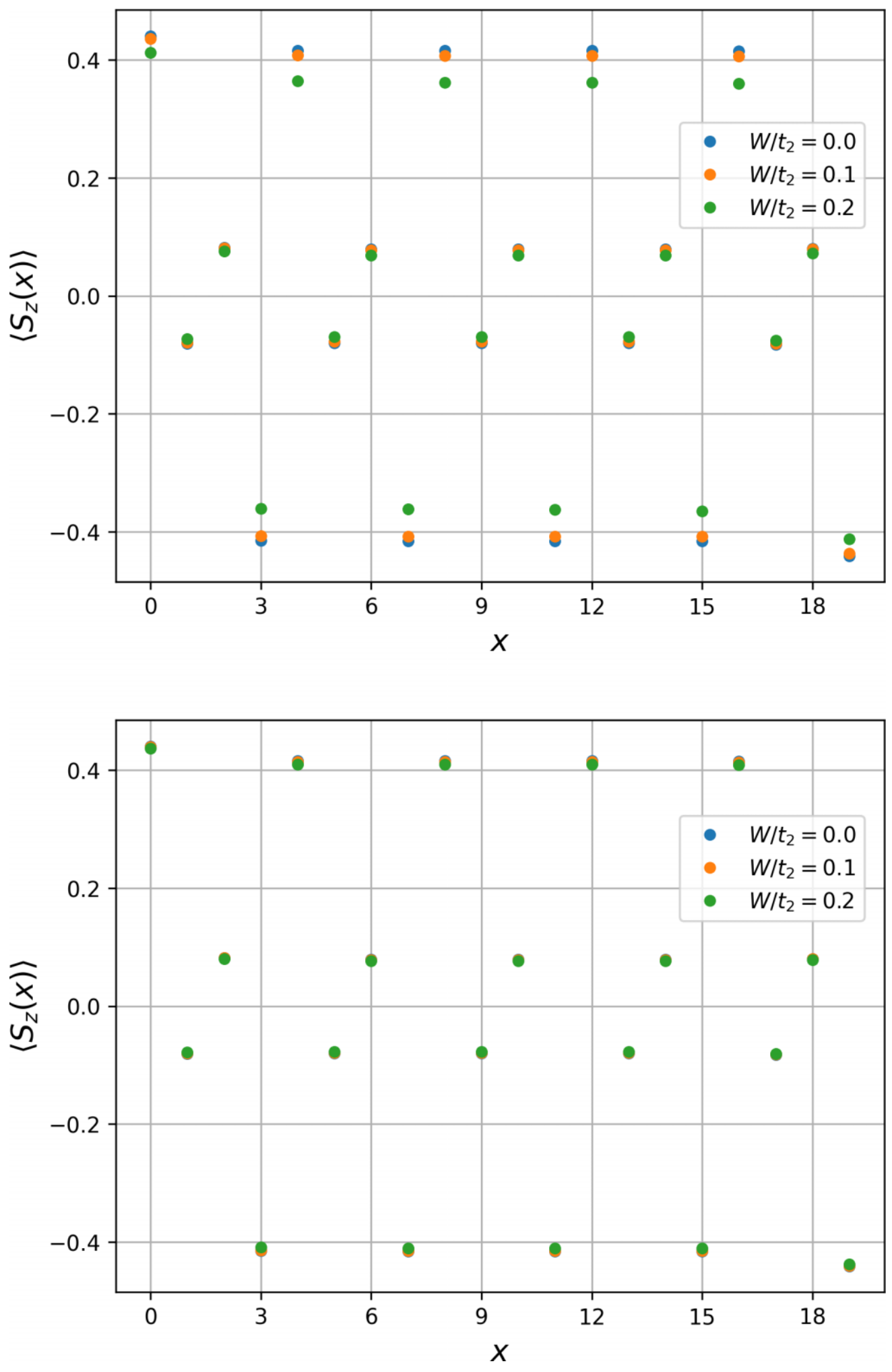

FIG. 7. Spin density distribution of the ground state in the SSH chain shown in the top of Fig. 5 with $N_x = 20$, $t_1/t_2 = 0.4$, $U = 0.5$, and $d_2/t_2 = (t_1/t_2)^2$. The results with select values of disorder amplitude $W$ are obtained from MF calculations for both (top) staggered on-site disorder and (bottom) random on-site disorder. Note that over 1000 disorder configurations are averaged to obtain the results in (bottom).

We now investigate the case of an odd SSH chain with defect engineering. For an odd SSH chain, the existence of the localized edge state is independent of the choice of the hopping parameters. Therefore, defect engineering can be used in all three cases shown in Figs. 2 and 8. Here, we choose $N_x = 21$ and $t_1/t_2 = 0.4$ which results in one unpaired spin center at the left end. The defects as shown in the top of Fig. 6 break the SSH chain into six blocks with a "4-4-4-4-4-1" configuration. With $U = 0.5$ as shown in the middle of Fig. 6, the ground state with total spin 1/2 consists of five paired spin singlets and one unpaired spin center at the right end. For excited states with larger total spins like $S_z = 3/2, 5/2, 7/2$, one, two, and more spin triplet pairings also arise, similar to the case of even SSH chain. As a result, the odd SSH chain with defect engineering can be used to realize the system with total half-integer spin hosting a number of connecting spin singlet and/or triplet pairings and one unpaired spin center. All these possible spin configurations in both even and odd SSH chain with defect engineering provide a promising platform for further applications using spin qubits based on spin singlets and/or spin triplets.

In addition to the chirality-preserving hopping disorder, we also take into account different types of disorder that can break chiral symmetry, such as a diagonal disorder coming from a fluctuating on-site chemical potential. The on-site disorder is introduced by the Hamiltonian

$$H_{\rm dis} = \sum_{x,\sigma} \delta_x n_{x,\sigma}. \tag{8}$$

For a staggered on-site disorder, we use $\delta_x = (-1)^x W$ [7,11]. A random on-site disorder $\delta_x$ has been chosen randomly from a uniform distribution on $[-W, W]$ [8–11].[2] Within the mean-field theory, we diagonalize the Hamiltonian $H_{\rm mf} + H_{\rm dis}$ self-consistently. In the noninteracting SSH model, the effects of on-site disorder have been well studied in previous works [7–11]. We find that the localized edge states still exist even in the absence of the chiral symmetry when the disorder is not too strong. In the present work, we focus on the interacting case and take the configuration in the top of Fig. 5 for example. With $U = 0.5$ and select values of $W$, the spin density distribution of the ground state is shown in Fig. 7 for both staggered and random on-site disorders. We find that the spin density distribution is quite robust against the weak disorder in the presence of interactions.

## V. EXPERIMENTAL FEASIBILITY

We now discuss the experimental feasibility of our work. The interacting many-body SSH chain has already been realized in experiments using silicon quantum dots [19] and artificial lattices made of Cs/InAs(111)A [20]. The intracell and intercell hopping strengths can be controlled in these experiments by varying the distance between quantum dots or artificial atoms. The control of these hopping strengths guarantees the system in a trivial or nontrivial topological phase. The hopping ratio we choose in our work is within the range of [0.18, 5.6] reported in these experiments.

[2]Note that we again normalize the disorder amplitude in a manner similar to $U$, i.e., "$W$"$= W/\max(t_1, t_2)$.

The on-site Hubbard interaction, on the other hand, is also tunable by varying the size of the quantum dot, as shown in Refs. [34–38]. A smaller radius results in a larger onsite $U$ and vice versa. One can also change $U$ by varying the number of Cs atoms in the artificial atom. It is expected that for asymptotically large values of $U$ the chain will have full antiferromagnetic order [39] and thus no longer have distinct spin centers, regardless of the presence of defects. Our numerical studies support this, as we can see that the relative heights of $\langle S_z^2(x)\rangle$ in the bottom of Figs. 5 and 6 become smaller for larger $U$. We emphasize that the relative height of $\langle S_z^2(x)\rangle$ remains nonzero even at very large $U = 10$, which suggests that the spin configurations are quite robust. Current experiments can easily probe systems with the values of $U$ used in this study.

As for defects, they are introduced by modifying the hoppings from $t_2$ to $d_2/t_2 = (t_1/t_2)^2$. This hopping ratio we use is also within the range accessible by experiments. The choice of the intercell hopping $d_2$ has no qualitative difference unless $d_2$ is close to or even larger than the intracell hopping $t_1$. As discussed in the last section, we also take into account the diagonal disorder possibly induced by the local fabrication defects and/or local field fluctuations in experimental realizations. The diagonal disorder would not be a relevant concern as long as the disorder is not too strong.

Currently the experimental size of these systems is rather limited, e.g., ten quantum dots and eight artificial atoms. Extending the size of the arrays will require more development.

## VI. SUMMARY

We have shown that localized edge states in both even and odd SSH chains together with the on-site Hubbard interaction can induce localized spin centers at the ends via exact diagonalization calculations, mean-field theory calculations, and also full QMC simulations. These localized spin centers are basic ingredients for engineering long SSH chains with multiple spin centers. This is done by introducing controllable defects with modified hoppings, giving rise to a long SSH chain that hosts a number of spin centers that form either spin-singlet or spin-triplet configurations. We find that these spin centers are quite robust for a wide range of interaction parameters and also robust against the weak on-site disorder. Furthermore, the constellation of parameters ($t_1/t_2$, $U$, and $N_x$) we used in this work fall within those accessible by current experiments. Because of their robustness, these spin-states open the possibility for making an array of spin qubits, thus providing a promising platform for future quantum applications such as many-body quantum simulations on the dynamics of spin excitations like magnons and triplons.

## ACKNOWLEDGMENTS

We thank Y. He, E. Berkowitz, J. Ostmeyer, L. Razmadze, and M. Rodekamp for fruitful discussions. This work was funded in part by the Deutsche Forschungsgemeinschaft (DFG, German Research Foundation) as part of the CRC 1639 NuMeriQS Project No. 511713970. We gratefully acknowledge the computing time granted by the JARA Vergabegremium and provided on the JARA Partition part of the supercomputer JURECA at Forschungszentrum Jülich [40]. The work of UGM was also supported in part by the CAS President's International Fellowship Initiative (PIFI) (Grant No. 2025PD0022).

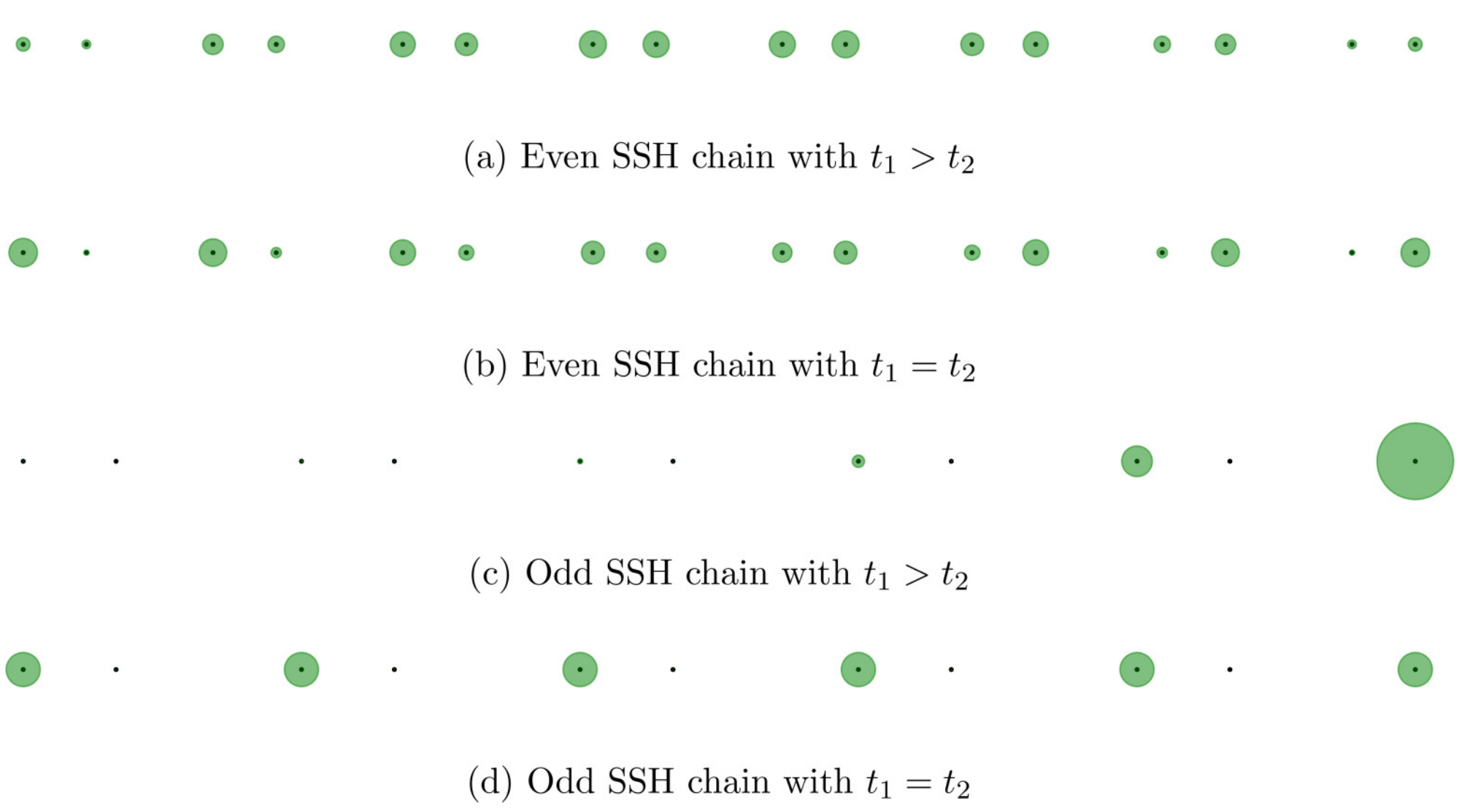

FIG. 8. (a), (b) In the SSH model with an even number of sites, the probability density of the state with lowest positive energy is shown as green circles. (c), (d) In the SSH model with an odd number of sites, the probability density of the state with exactly zero energy is plotted as green circles. When $t_1 = t_2$, the zero-energy state is evenly localized at all A sites.

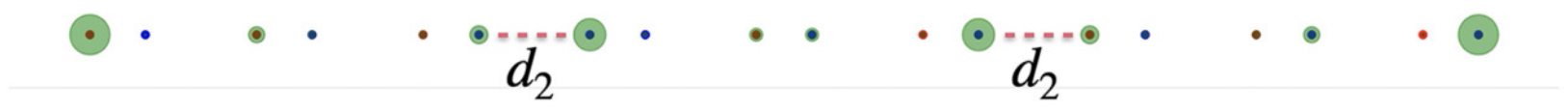

(a) Defects $t_2 \to d_2$, $U = 0$

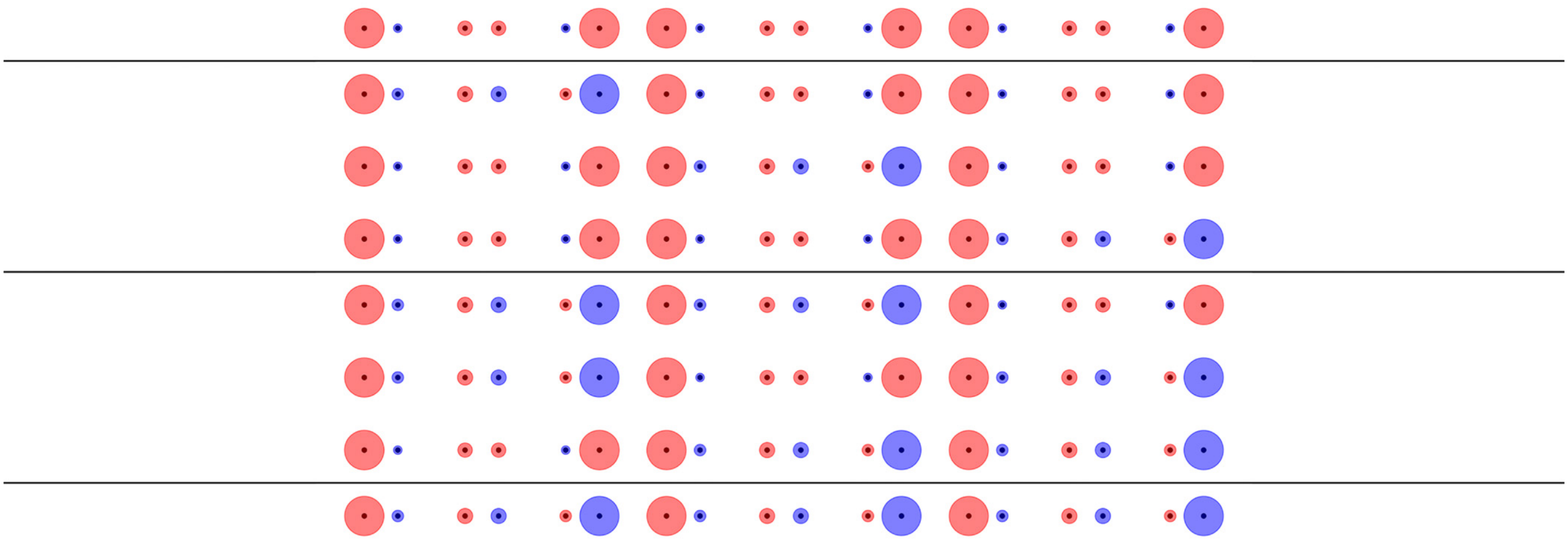

(b) Defects $t_2 \to d_2$, $S_z = 0, 1, 2, 3$ from bottom to top

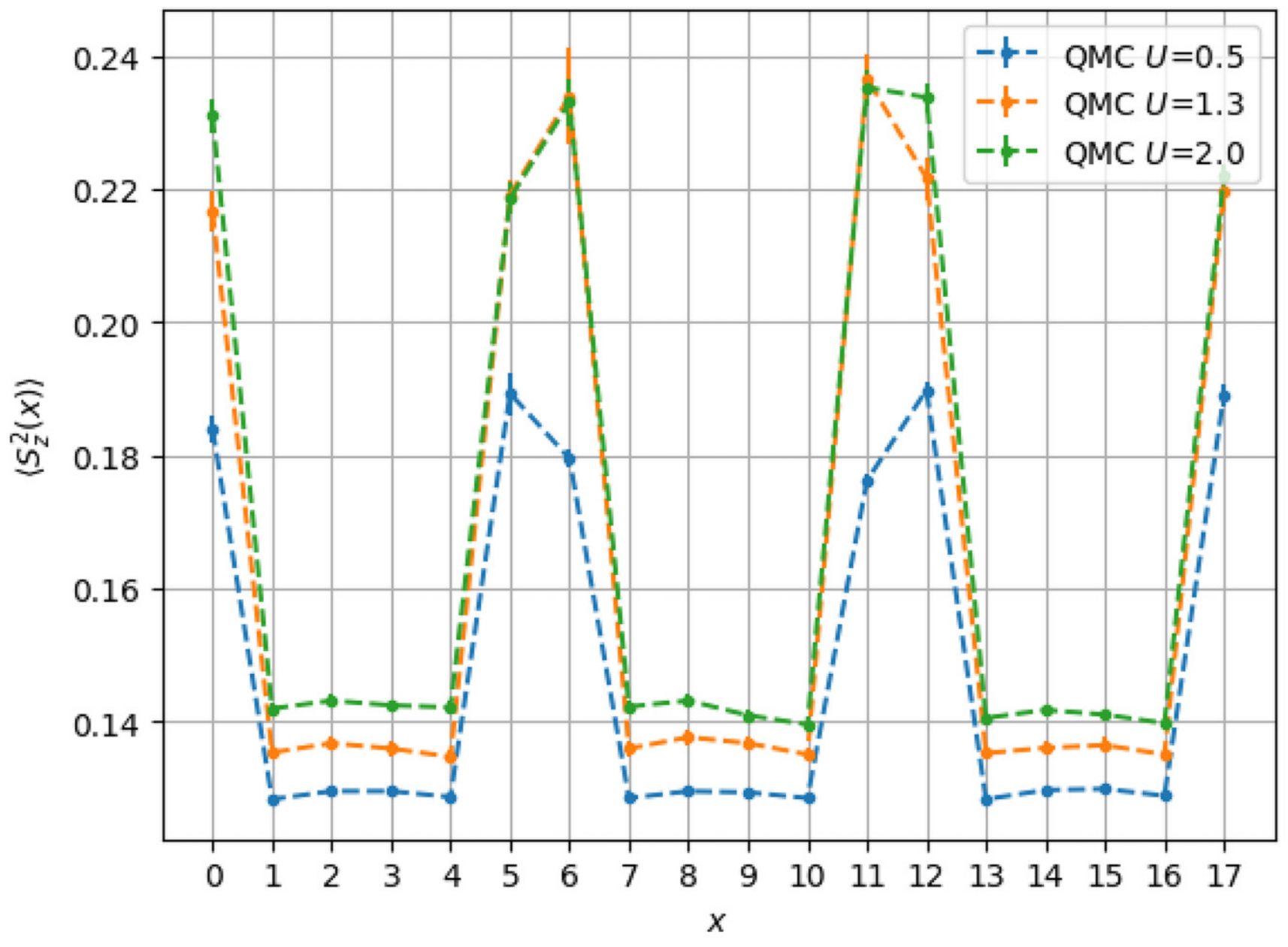

(c) $\langle S_z^2(x) \rangle$

FIG. 9. Defect engineering the even SSH chain with $N_x = 18$ and $t_1/t_2 = 0.4$ in the presence of Hubbard interaction $U$. The defects are introduced by modifying the hoppings $t_2 \to d_2$. (a) The probability density of state with exactly zero energy is plotted as green circles. (b) Spin density distribution of the states obtained via MF at $U = 0.5$ is shown as red (spin up) and blue (spin down) circles. (c) $\langle S_z^2(x) \rangle$ of the ground state obtained from QMC results with select values of $U$ and $\beta = 10$. $S_z$ is the total spin. $d_2/t_2 = (t_1/t_2)^3$.

## DATA AVAILABILITY

The data that support the findings of this article are not publicly available because of legal restrictions preventing unrestricted public distribution. The data are available from the authors upon reasonable request.

## APPENDIX A: NONINTERACTING EVEN AND ODD SSH CHAIN

In finite SSH chain with chiral symmetry, the zero-energy states have well defined chirality $+1$ or $-1$. The odd SSH chain with one unpaired A (B) at the end only has

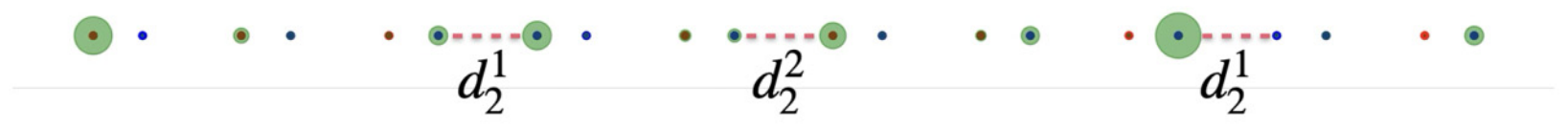

(a) Defects $t_2 \to d_2$, $U = 0$

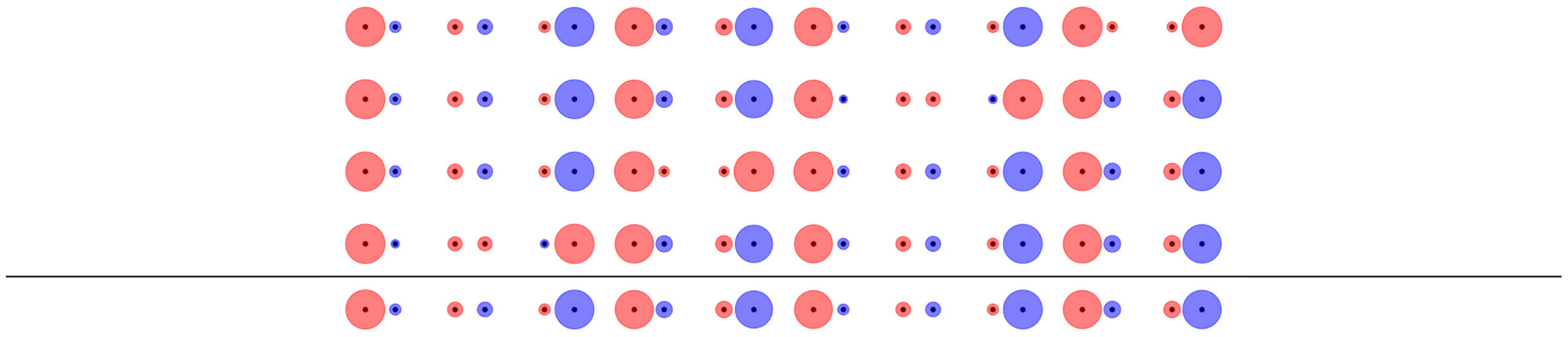

(b) Defects $t_2 \to d_2$, $S_z = 0, 1$ from bottom to top

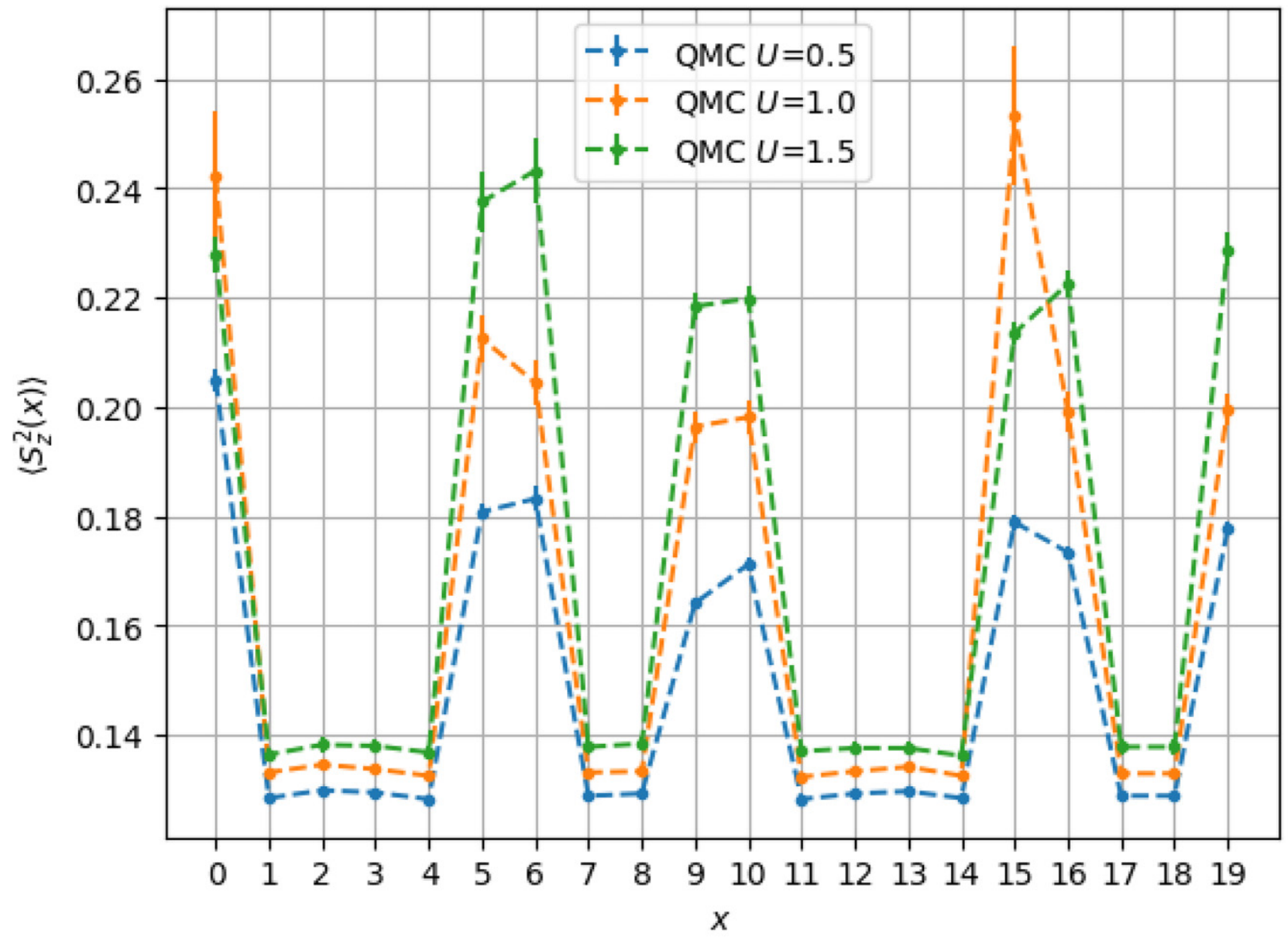

(c) $\langle S_z^2(x)\rangle$

FIG. 10. Defect engineering the even SSH model with $N_x = 20$ and $t_1/t_2 = 0.4$ in the presence of Hubbard interaction $U$. The defects are introduced by modifying the hoppings $t_2 \to d_2$. (a) The probability density of state with exactly zero energy is plotted as green circles. (b) Spin density distribution of the states obtained via MF at $U = 0.5$ is shown as red (spin up) and blue (spin down) circles. (c) $\langle S_z^2(x)\rangle$ of the ground state obtained from QMC results with select values of $U$ and $\beta = 10$. $d_2^1/t_2 = (t_1/t_2)^3$ and $d_2^2/t_2 = (t_1/t_2)^2$.

zero-energy state with chirality $+1$ $(-1)$. The chirality $+1$ $(-1)$ determines the wavefunction of all B (A) sites to be zero. Here, we take the odd SSH chain with one unpaired A at the end, for example, to derive the detailed wavefunctions of the zero-energy states. As mentioned previously, the wavefunctions of all B sites vanish and therefore we only need to determine the components of A sites. From $H_{\rm SSH}\Phi = 0$, we obtain $t_1\phi_{i,{\rm A}} + t_2\phi_{i+1,{\rm A}} = 0$ with $\phi_{i,{\rm A}}$ being the $i$th component of A sites. There are three possible situations depending on the hopping ratio $t_1/t_2$. (i) When $t_1 = t_2$, we have $\phi_{i,{\rm A}} = (-1)^{i-1}\phi_{1,{\rm A}}$ with all A sites evenly distributed. (ii) When $t_1 < t_2$, to obtain stable state, zero-energy state has to be localized at the left end with $\phi_{i,{\rm A}} = (-t_1/t_2)^{i-1}\phi_{1,{\rm A}}$ and the decay length is $\xi = \frac{1}{\ln(t_2/t_1)}$. (iii) When $t_1 > t_2$, the stable zero-energy state has to be localized at the right end with $\phi_{i,{\rm A}} = (-t_2/t_1)^{N_A - i}\phi_{N_A,{\rm A}}$, $N_A$ is the number of all A sites, and the decay length is $\xi = \frac{1}{\ln(t_1/t_2)}$. Similarly, for even SSH chain in the topological phase $t_1 < t_2$, the decay length is also given by $\xi = \frac{1}{\ln(t_2/t_1)}$. For completeness, we also show in Fig. 8 the probability density of the states with exactly zero energy or lowest positive energy in the cases of $t_1 > t_2$ and $t_1 = t_2$ for both even and odd SSH chains.

## APPENDIX B: USING DEFECTS TO ENGINEER SPIN CENTERS

Defects can be introduced by modifying the intracell and/or intercell hoppings, e.g., $t_1 \to d_1$ and/or $t_2 \to d_2$, or changing the on-site energy of the sites. Here, we focus on the defects with respect to the hoppings since they do not break the chiral symmetry. We take the odd SSH chain with one unpaired A at the end for example. In the case of $t_1 < t_2$, there exists one zero-energy state localized at the left end. By introducing a defect between $m$th A and $m$th B with hopping $t_1 \to d_1$, $\phi_{m+1,\mathrm{A}}/\phi_{m,\mathrm{A}}$ changes from $-t_1/t_2$ to $-d_1/t_2$. When $d_1 < t_1$, the localization at site $m+1$th A will be suppressed. However, when $d_1 > t_1$, the localization at site $m+1$-A will be enhanced. When $d_1$ is large enough so that $\phi_{m+1,\mathrm{A}}$ is comparable with $\phi_{1,\mathrm{A}}$, two localized centers are created. When $d_1$ is further increased such that $\phi_{m+1,\mathrm{A}}$ is even larger than $\phi_{1,\mathrm{A}}$, a different localized center arises at site $m+1$-A. On the other hand, by introducing a defect between $m+1$th A and $m$th B with hopping $t_2 \to d_2$, $\phi_{m+1,\mathrm{A}}/\phi_{m,\mathrm{A}}$ changes from $-t_1/t_2$ to $-t_1/d_2$. When $d_2 > t_2$, the localization at site $m+1$th A will be suppressed. However, when $d_2 < t_2$, the localization at site $m+1$th A will be enhanced. When $d_2$ is small enough so that $\phi_{m+1,\mathrm{A}}$ is comparable or even larger than $\phi_{1,\mathrm{A}}$, two localized centers or one different localized center can be created. Therefore, by engineering the defects via increasing $d_1/t_1$ and/or decreasing $d_2/t_2$ to enhance the localizations, one can easily realize all kinds of localized centers with controllable numbers and the site positions in the odd SSH chain. Such defect engineering can be also extended to the even SSH chain.

For the even SSH chain in the topological phase, in addition to the configuration shown in Fig. 5, defect engineering is also applied to other configurations such as 18 sites with "6-6-6" in Fig. 9 and 20 sites with "6-4-6-4" in Fig. 10. Both of them show a chain connected by a number of spin singlets and/or spin triplets.

[1] W. P. Su, J. R. Schrieffer, and A. J. Heeger, Solitons in polyacetylene, Phys. Rev. Lett. **42**, 1698 (1979).

[2] M. Z. Hasan and C. L. Kane, *Colloquium*: Topological insulators, Rev. Mod. Phys. **82**, 3045 (2010).

[3] X.-L. Qi and S.-C. Zhang, Topological insulators and superconductors, Rev. Mod. Phys. **83**, 1057 (2011).

[4] L. Barbiero, L. Santos, and N. Goldman, Quenched dynamics and spin-charge separation in an interacting topological lattice, Phys. Rev. B **97**, 201115(R) (2018).

[5] M. Bahari and M. V. Hosseini, Topological properties of a generalized spin–orbit-coupled Su–Schrieffer–Heeger model, Physica E **119**, 113973 (2020).

[6] D. Sticlet, L. Seabra, F. Pollmann, and J. Cayssol, From fractionally charged solitons to Majorana bound states in a one-dimensional interacting model, Phys. Rev. B **89**, 115430 (2014).

[7] L. Qi, G.-L. Wang, S. Liu, S. Zhang, and H.-F. Wang, Engineering the topological state transfer and topological beam splitter in an even-sized Su-Schrieffer-Heeger chain, Phys. Rev. A **102**, 022404 (2020).

[8] M. Scollon and M. P. Kennett, Persistence of chirality in the Su-Schrieffer-Heeger model in the presence of on-site disorder, Phys. Rev. B **101**, 144204 (2020).

[9] E. G. Cinnirella, A. Nava, G. Campagnano, and D. Giuliano, Fate of high winding number topological phases in the disordered extended Su-Schrieffer-Heeger model, Phys. Rev. B **109**, 035114 (2024).

[10] M. P. Estarellas, I. D'Amico, and T. P. Spiller, Topologically protected localised states in spin chains, Sci. Rep. **7**, 42904 (2017).

[11] S. Mandal and S. Kar, Topological solitons in a Su-Schrieffer-Heeger chain with periodic hopping modulation, domain wall, and disorder, Phys. Rev. B **109**, 195124 (2024).

[12] K. Sone, M. Ezawa, Z. Gong, T. Sawada, N. Yoshioka, and T. Sagawa, Transition from the topological to the chaotic in the nonlinear Su–Schrieffer–Heeger model, Nat. Commun. **16**, 422 (2025).

[13] W. Cai, J. Han, F. Mei, Y. Xu, Y. Ma, X. Li, H. Wang, Y. P. Song, Z.-Y. Xue, Z.-q. Yin, S. Jia, and L. Sun, Observation of topological magnon insulator states in a superconducting circuit, Phys. Rev. Lett. **123**, 080501 (2019).

[14] E. Kim, X. Zhang, V. S. Ferreira, J. Banker, J. K. Iverson, A. Sipahigil, M. Bello, A. González-Tudela, M. Mirhosseini, and O. Painter, Quantum electrodynamics in a topological waveguide, Phys. Rev. X **11**, 011015 (2021).

[15] E. J. Meier, F. A. An, and B. Gadway, Observation of the topological soliton state in the Su-Schrieffer-Heeger model, Nat. Commun. **7**, 13986 (2016).

[16] P. St-Jean, V. Goblot, E. Galopin, A. Lemaître, T. Ozawa, L. Le Gratiet, I. Sagnes, J. Bloch, and A. Amo, Lasing in topological edge states of a one-dimensional lattice, Nat. Photon. **11**, 651 (2017).

[17] S. de Léséleuc, V. Lienhard, P. Scholl, D. Barredo, S. Weber, N. Lang, H. P. Büchler, T. Lahaye, and A. Browaeys, Observation of a symmetry-protected topological phase of interacting bosons with Rydberg atoms, Science **365**, 775 (2019).

[18] S. Kruk, A. Slobozhanyuk, D. Denkova, A. Poddubny, I. Kravchenko, A. Miroshnichenko, D. Neshev, and Y. Kivshar, Edge states and topological phase transitions in chains of dielectric nanoparticles, Small **13**, 1603190 (2017).

[19] M. Kiczynski, S. K. Gorman, H. Geng, M. B. Donnelly, Y. Chung, Y. He, J. G. Keizer, and M. Y. Simmons, Engineering topological states in atom-based semiconductor quantum dots, Nature (London) **606**, 694 (2022).

[20] R. A. M. Ligthart, M. A. J. Herrera, A. C. H. Visser, A. Vlasblom, D. Bercioux, and I. Swart, Wannier center spectroscopy to identify boundary-obstructed topological insulators, Phys. Rev. Res. **7**, 012076 (2025).

[21] O. Katz, L. Feng, D. Porras, and C. Monroe, Floquet control of interactions and edge states in a programmable quantum simulator, Nat. Commun. **16**, 8815 (2025).

[22] P. C. Fariña, D. Jirovec, X. Zhang, E. Morozova, S. D. Oosterhout, S. Reale, T.-K. Hsiao, G. Scappucci, M. Veldhorst, and L. M. Vandersypen, Site-resolved magnon and triplon

dynamics on a programmable quantum dot spin ladder, arXiv:2506.08663.

[23] A. P. Schnyder, S. Ryu, A. Furusaki, and A. W. W. Ludwig, Classification of topological insulators and superconductors in three spatial dimensions, Phys. Rev. B **78**, 195125 (2008).

[24] C.-K. Chiu, J. C. Y. Teo, A. P. Schnyder, and S. Ryu, Classification of topological quantum matter with symmetries, Rev. Mod. Phys. **88**, 035005 (2016).

[25] M. Sato, Y. Tanaka, K. Yada, and T. Yokoyama, Topology of Andreev bound states with flat dispersion, Phys. Rev. B **83**, 224511 (2011).

[26] T. Luu and T. A. Lähde, Quantum Monte Carlo calculations for carbon nanotubes, Phys. Rev. B **93**, 155106 (2016).

[27] J. Ostmeyer, E. Berkowitz, S. Krieg, T. A. Lähde, T. Luu, and C. Urbach, Semimetal–Mott insulator quantum phase transition of the Hubbard model on the honeycomb lattice, Phys. Rev. B **102**, 245105 (2020).

[28] J. Ostmeyer, E. Berkowitz, S. Krieg, T. A. Lähde, T. Luu, and C. Urbach, Antiferromagnetic character of the quantum phase transition in the Hubbard model on the honeycomb lattice, Phys. Rev. B **104**, 155142 (2021).

[29] T. Luu, U.-G. Meißner, and L. Razmadze, Localization of electronic states in hybrid nanoribbons in the nonperturbative regime, Phys. Rev. B **106**, 195422 (2022).

[30] J. Ostmeyer, L. Razmadze, E. Berkowitz, T. Luu, and U.-G. Meißner, Effective theory for graphene nanoribbons with junctions, Phys. Rev. B **109**, 195135 (2024).

[31] U. Las Heras, A. Mezzacapo, L. Lamata, S. Filipp, A. Wallraff, and E. Solano, Digital quantum simulation of spin systems in superconducting circuits, Phys. Rev. Lett. **112**, 200501 (2014).

[32] K. De Greve, L. Yu, P. L. McMahon, J. S. Pelc, C. M. Natarajan, N. Y. Kim, E. Abe, S. Maier, C. Schneider, M. Kamp, S. Höfling, R. H. Hadfield, M. M. Forchel, and Y. Yamamoto, Quantum-dot spin–photon entanglement via frequency downconversion to telecom wavelength, Nature (London) **491**, 421 (2012).

[33] J. R. Maze, P. L. Stanwix, J. S. Hodges, S. Hong, J. M. Taylor, P. Cappellaro, L. Jiang, M. V. G. Dutt, E. Togan, A. S. Zibrov, A. Yacoby, R. L. Walsworth, and M. D. Lukin, Nanoscale magnetic sensing with an individual electronic spin in diamond, Nature (London) **455**, 644 (2008).

[34] L. Wang, K. Shen, B. Y. Sun, and M. W. Wu, Singlet-triplet relaxation in multivalley silicon single quantum dots, Phys. Rev. B **81**, 235326 (2010).

[35] L. Wang and M. W. Wu, Singlet-triplet relaxation in SiGe/Si/SiGe double quantum dots, J. Appl. Phys. **110**, 043716 (2011).

[36] Z. Liu, L. Wang, and K. Shen, Energy spectra of three electrons in Si/SiGe single and vertically coupled double quantum dots, Phys. Rev. B **85**, 045311 (2012).

[37] Y. Ren, L. Wang, Z. Liu, and M. Wu, Energy spectra of three electrons in SiGe/Si/SiGe laterally coupled triple quantum dots, Physica E **63**, 329 (2014).

[38] G. Hu, W. W. Huang, R. Cai, L. Wang, C. H. Yang, G. Cao, X. Xue, P. Huang, and Y. He, Single-electron spin qubits in silicon for quantum computing, Intell. Comput. **4**, 0115 (2025).

[39] P. W. Anderson, New approach to the theory of superexchange interactions, Phys. Rev. **115**, 2 (1959).

[40] Jülich Supercomputing Centre, JURECA: Data centric and booster modules implementing the modular supercomputing architecture at Jülich Supercomputing Centre, J. Large Scale Res. Facil. **7**, A182 (2021).